\documentclass[
  aps,prd,reprint,superscriptaddress,nofootinbib,amsmath,amssymb
]{revtex4-2}

\usepackage[T1]{fontenc}
\usepackage[utf8]{inputenc}
\usepackage{graphicx}
\usepackage{bm}
\usepackage{booktabs}
\usepackage{xcolor}
\usepackage[hidelinks]{hyperref}
\usepackage{orcidlink}
\usepackage{comment}
\usepackage{ulem}

\newcommand{\Nc}{N_c}
\newcommand{\dd}{\mathrm{d}}
\newcommand{\order}[1]{\mathcal O\!\left(#1\right)}
\newcommand{\Eden}{\mathcal E}

\newcommand{\orcidauthorA}{0009-0000-5341-562X} 
\newcommand{\orcidauthorB}{0000-0003-1347-0821} 
\newcommand{\orcidauthorC}{0000-0002-0016-6612} 

\begin{document}

\title{Construction of a conserved energy-momentum tensor of the Skyrmion coupled to additional dynamical degrees in the large-$\Nc$ expansion}

\author{Mitsuru Tanaka\,\orcidlink{\orcidauthorA}}
\email[]{tanaka@hken.phys.nagoya-u.ac.jp}
\affiliation{Department of Physics, Nagoya University, Nagoya, 464-8602, Japan}

\author{Yasuhiro Yamaguchi\,\orcidlink{\orcidauthorB}}
\email[]{yyamaguchi@tmu.ac.jp}
\affiliation{Department of Physics, Tokyo Metropolitan University, Hachioji, 191-0397, Japan}
\affiliation{Department of Physics, Nagoya University, Nagoya, 464-8602, Japan}
\affiliation{Kobayashi-Maskawa Institute for the Origin of Particles and the Universe, Nagoya University, Nagoya, 464-8602, Japan}

\author{Masayasu Harada\,\orcidlink{\orcidauthorC}}
\email{harada@hken.phys.nagoya-u.ac.jp}
\affiliation{Kobayashi-Maskawa Institute for the Origin of Particles and the Universe, Nagoya University, Nagoya, 464-8602, Japan}
\affiliation{Department of Physics, Nagoya University, Nagoya, 464-8602, Japan}
\affiliation{Advanced Science Research Center, Japan Atomic Energy Agency, Tokai, 319-1195, Japan}

\begin{abstract}
We formulate an order-by-order construction of a conserved energy-momentum tensor (EMT) in the large-$\Nc$ expansion of the model where the Skyrmion is
coupled to additional dynamical fields.
We show that, expanding the equations of motion and the EMT in powers of $1/N_c$, the inclusion of the backreaction to the Skyrmion from the additional fields is essential for achieving the conservation law of the EMT, which is contrasted with the fixed-background approximation where the effects of backreaction are neglected.
We apply the construction to the model for hyperons based on the Callan-Klebanov bound-state approach, where the antikaon is bound to the Skyrmion, to obtain the EMT up to next-to-leading order of the $1/N_c$ expansion.
We explicitly show that the divergence of the EMT from the backreaction for the Skyrmion exactly cancels that of the kaon to achieve the conservation of the total EMT.
We find that the conserved EMT gives a negative hyperon D-term whose magnitude is smaller than that of the nucleon D-term and radii of gravitational form factors for the hyperon smaller than those for the nucleon, which implies that the hyperon is more compact than the nucleon.
\end{abstract}

\maketitle
\renewcommand{\thefootnote}{\#\arabic{footnote}}

\section{Introduction}\label{sec:introduction}

Gravitational form factors (GFFs) are defined by the hadronic matrix elements of the energy-momentum tensor (EMT) and they provide information about the energy, momentum, and angular momentum of hadrons~\cite{Polyakov:2018zvc}.
In particular, the D-term probes the spatial components of the EMT and has attracted considerable attention as a global property complementary to the mass and spin of hadrons~\cite{Polyakov:2018zvc}.
Although direct measurements through gravitational interactions are practically impossible, GFFs can be accessed indirectly through hard exclusive processes~\cite{Ji:1996ek,Polyakov:2002yz}.
Indeed, the proton's quark D-term was analyzed experimentally for the first time in 2018~\cite{Burkert:2018bqq}, and the Electron-Ion Collider is expected to provide further constraints on quark and gluon GFFs~\cite{Accardi:2012qut,AbdulKhalek:2021gbh}.

These developments motivate theoretical calculations of GFFs such as lattice QCD~\cite{Yang:2014xsa,Shanahan:2018pib,Alexandrou:2019ali,ExtendedTwistedMass:2021rdx,Pefkou:2021fni,Hackett:2023rif,Hackett:2023nkr,Wang:2024lrm,Abbott:2025irb}, perturbative QCD~\cite{Tong:2021ctu,Tong:2022zax,Tanaka:2022wzy}, chiral effective field theory~\cite{Donoghue:1991qv,Kubis:1999db,Belitsky:2002jp,Alharazin:2020yjv,Epelbaum:2021ahi,Gegelia:2021wnj,Alharazin:2022wjj,Alharazin:2023uhr,Alharazin:2023zzc}, sum rules~\cite{Anikin:2019kwi}, and dispersion relations~\cite{Cao:2024zlf,Cao:2025dkv}, as well as various other approaches~\cite{Chakrabarti:2015lba,Neubelt:2019sou,Freese:2019bhb,Won:2022cyy,Fujita:2022jus,Yao:2024ixu,Liu:2025qja}.
The Skyrme model~\cite{Skyrme:1961vq,Skyrme:1962vh,Adkins:1983ya} provides a practical framework for organizing baryon properties in the large-$\Nc$ expansion and has been extensively employed to study GFFs of nucleons~\cite{Cebulla:2007ei,Fukushima:2026wwc}, $\Delta$ baryons~\cite{Kim:2020lrs}, nuclei~\cite{GarciaMartin-Caro:2023klo,GarciaMartin-Caro:2023toa}, and so on~\cite{Jung:2013bya,Kim:2012ts,Jung:2014jja,Fujii:2025aip,Tanaka:2025pny,Fukushima:2025jah,Fukushima:2026wwc}.

Numerous extensions of the Skyrme model have been developed by introducing additional fields and interactions~\cite{Meissner:1986ka,Meissner:1986js,Gomm:1986fe, Lee:2003eg,Park:2008zg,Tsushima:1992nqq}, including the kaon-soliton bound-state approach proposed by Callan and Klebanov~\cite{Callan:1985hy,Callan:1987xt} and the heavy-meson-soliton bound-state approach~\cite{Jenkins:1992zx,Momen:1993ax,Gupta:1993kd,Schechter:1994ip,Schechter:1995vr,Harada:1997bk,Oh:1994vd,Oh:1994yv,Oh:1995ey}.
In the following we call these models the ``bound-Skyrmion models''.
A commonly employed approximation in these extensions is to first solve the equation of motion (EOM) for the pion field derived from the original action and then solve the EOMs for the additional fields while assuming that the pion field remains unaffected~\cite{Callan:1985hy,Callan:1987xt}.
In this paper, we call this treatment the ``fixed-background approximation''.
Although this approximation is sufficient to evaluate the mass, its use in the calculation of GFFs requires additional care.
Actually, the fixed-background approximation does not, in general, yield a conserved total EMT, which results from the inconsistency with the large-$\Nc$ expansion.
This is because the backreaction from the additional sector to the pion field is neglected in the fixed-background approximation, even though its contribution to the EMT is of the same order in the large-$\Nc$ expansion as that from the additional sector.
Throughout this paper, we use ``backreaction'' to mean the change in the pion field caused by its coupling to the additional dynamical field.

Motivated by the growing interest in GFFs and the above limitations of the fixed-background approximation in the bound-Skyrmion models, the aim of this work is to establish a systematic method for constructing a conserved EMT in the models consistently with the large-$\Nc$ expansion.
After showing the general construction, we apply it to the model based on the ``Callan-Klebanov (CK) bound-state approach''~\cite{Callan:1985hy,Callan:1987xt}, in which the hyperon is described as a bound state of an antikaon and a Skyrmion.
We construct the EMT of hyperon through next-to-leading order in the large-$N_c$ expansion and the GFFs. 

The paper is organized as follows.
Section~\ref{sec:gffs} introduces the gravitational form factors and their extraction from the static EMT.
Section~\ref{sec:skyrme} presents the bound-Skyrmion model and the general construction of a conserved EMT in the large-$\Nc$ expansion.
Section~\ref{sec:model} applies the method to a model for hyperons.
Section~\ref{sec:results} presents numerical results for the GFFs, focusing on the effects of backreaction and the differences between the hyperon and the nucleon.
Section~\ref{sec:conclusions} summarizes our findings and discusses future extensions.

\section{Energy-momentum tensor and gravitational form factors}\label{sec:gffs}

\subsection{Gravitational form factors}

The matrix element of the EMT for a spin-$1/2$ baryon of mass $M$ is parametrized as~\cite{Polyakov:2018zvc}
\begin{align}
  &\langle p',s'|T^{\mu\nu}(0)|p,s\rangle \notag\\
  &\quad=\bar u'(p',s') \Biggl[A(t)\frac{P^\mu P^\nu}{M} +J(t)\frac{i(P^\mu\sigma^{\nu\rho}+P^\nu\sigma^{\mu\rho})\Delta_\rho}{2M}
    \notag\\
  &\qquad+D(t)\frac{\Delta^\mu\Delta^\nu-g^{\mu\nu}\Delta^2}{4M}
    \Biggr]u(p,s) .
  \label{eq:gff-parametrization}
\end{align}
Here $P=(p'+p)/2$, $\Delta=p'-p$, $t=\Delta^2$, and $\sigma^{\mu\nu}=i[\gamma^\mu,\gamma^\nu]/2$.
We normalize the spinors by $\bar u(p,s)u(p,s)=2M$ and the states by
$\langle p',s'|p,s\rangle=2p^0(2\pi)^3
\delta^{(3)}(\bm p'-\bm p)\delta_{s's}$.
We use the metric $g_{\mu\nu}=\operatorname{diag}(1,-1,-1,-1)$.
Momentum and angular-momentum conservation imply $A(0)=1$ and $J(0)=1/2$, whereas the D-term, $D(0)$, is not fixed by conservation laws.

The scalar (trace) form factor $\theta(t)$ is defined directly through the trace of the EMT:
\begin{align}
   \langle p',s'|T^\mu{}_{\mu}(0)|p,s\rangle
    &=\bar u'(p',s')\,M \theta(t) u(p,s) .
    \label{eq:scalar-gff-definition}
\end{align}
Contracting Eq.~\eqref{eq:gff-parametrization} with $g_{\mu\nu}$, we obtain the following relation:
\begin{align}
  \theta(t)&=A(t) - \frac{t}{4M^2}\bigl[A(t)-2J(t)+3D(t)\bigr].
    \label{eq:scalar-gff}
\end{align}
When $D(0)$ is finite, the above equation together with $A(0)=1$ and $J(0)=1/2$ implies that $\theta(0)=1$.

\subsection{Static EMT and conservation}

In the Breit frame, where $P=(E,\bm0)$, $\Delta=(0,\bm q)$, $t=-q^2$, and $E=\sqrt{M^2+q^2/4}$, we define the static EMT as the Fourier transform of the matrix element~\cite{Polyakov:2018zvc}:
\begin{align}
  T_{\mathrm{stat}}^{\mu\nu}(\bm r)
    &=\int\frac{\dd^3q}{(2\pi)^3}\,e^{-i\bm q\cdot\bm r}
    \frac{\langle p',s|T^{\mu\nu}(0)|p,s\rangle}{2E} ,
  \label{eq:breit-emt}
\end{align}
and the energy density as
\begin{align}
  \Eden(r)=T_{\mathrm{stat}}^{00}(\bm r), \label{eq:energy-density}
\end{align}
where $r=|\bm r|$.
For the spin-independent sector of a spin-$1/2$ baryon, it is convenient
to decompose the spatial EMT into its trace and traceless parts:
\begin{equation}
  T_{\mathrm{stat}}^{ij}(\bm r)
    =\left(\frac{r^i r^j}{r^2}-\frac{\delta^{ij}}3\right)s(r)
      +\delta^{ij}p(r).
  \label{eq:spatial-emt-densities}
\end{equation}
Here $p(r)$ and $s(r)$ are the coefficients of the spatial trace and traceless parts of the EMT, respectively.
The interpretation of $p(r)$ and $s(r)$, which could be regarded as pressure and shear force in the literature, has been discussed from different perspectives~\cite{Polyakov:2018zvc,Lorce:2025oot,Mejia:2025oip,Freese:2024rkr,Ji:2025qax}.
Here, we use these functions as coefficients of the decomposition of the spatial EMT and as intermediate functions for extracting the GFFs and checking the conservation of the EMT, without relying on a particular interpretation.

For a conserved EMT, $\partial_iT_{\mathrm{stat}}^{ij}=0$\footnote{From Eq.~\eqref{eq:gff-parametrization}, $\partial_iT_{\mathrm{stat}}^{i0}=0$ is identically satisfied in the Breit frame.} gives~\cite{Polyakov:2018zvc,Cebulla:2007ei}
\begin{equation}
  p'(r)+\frac23s'(r)+\frac{2s(r)}r=0.
  \label{eq:local-stress-conservation}
\end{equation}
With regularity at the origin and sufficiently fast decay at infinity,
this implies the von Laue condition
\begin{equation}
  \int\dd^3r\,p(r)=4\pi\int_0^\infty\dd r\,r^2p(r)=0.
  \label{eq:von-laue}
\end{equation}
We would like to stress that the functions $p(r)$ and $s(r)$ determined from Eq.~(\ref{eq:spatial-emt-densities}) should satisfy these conservation laws~\eqref{eq:local-stress-conservation} and \eqref{eq:von-laue}.

\subsection{GFFs and mean-square radii}
\label{sec:static-emt-gffs}

For a conserved EMT, $D(t)$ can be extracted equivalently from either coefficient, $p(r)$ or $s(r)$~\mbox{\cite{Polyakov:2018zvc,Cebulla:2007ei}}:
\begin{align}
  D(-q^2)&=-\frac{6M}{q^2}\int\dd^3r\,
    [j_0(qr)-1]p(r) \notag \\
  &=-\frac{4M}{q^2}\int\dd^3r\,j_2(qr)s(r),
    \label{eq:d-from-stress}
\end{align}
where $j_\ell$ denotes a spherical Bessel function.
The conservation law~\eqref{eq:local-stress-conservation} guarantees that the expression for $D(t)$ derived from $p(r)$ is identical to that from $s(r)$.
By expanding the spherical Bessel functions and taking the limit $q\to 0$, we obtain the D-term represented by the second spatial moments of $p(r)$ and $s(r)$: 
\begin{align}
  D(0)&=M\int\dd^3r\,r^2p(r)\notag\\
      &=-\frac{4M}{15}\int\dd^3r\,r^2s(r).
  \label{eq:d-term}
\end{align}
In particular, $D(0)/M$ isolates the spatial EMT moment when comparing states with different masses.

We can extract $A(t)$ from the Fourier transform of the energy density $\Eden(r) = T^{00}_{\rm stat} (\bm r)$, which is related to the GFFs by
\begin{align}
  & \frac1M\int\dd^3r\,j_0(r\sqrt{-t})\Eden(r) \notag \\ 
  & = A(t)-\frac{t}{4M^2}\bigl[A(t)-2J(t)+D(t)\bigr].
  \label{eq:energy-gff-exact}
\end{align}
Here $M=\int\dd^3r\,\Eden(r)$ is the baryon mass.
At fixed $t=\order{\Nc^0}$, the mass and form factors are counted as $M=\order{\Nc}$, $A,J=\order{\Nc^0}$, and $D=\order{\Nc^2}$, which allows us to neglect $t[A(t)-2J(t)]/4M^2$ while retaining $tD(t)/4M^2$~\cite{Cebulla:2007ei}.
Therefore, we can approximate Eq.~(\ref{eq:energy-gff-exact}) as
\begin{align}
  & \frac1M\int\dd^3r\,j_0(r\sqrt{-t})\Eden(r) = A(t) - \frac{t}{4M^2} D(t).
  \label{eq:a-from-energy}
\end{align}
By substituting Eq.~(\ref{eq:scalar-gff}) into the above expression, the energy density is related to the scalar form factor as
\begin{align}
 \frac1M\int\dd^3r\,j_0(r\sqrt{-t})\Eden(r) = \theta(t) + \frac{t}{2M^2} D(t).
  \label{eq:trace-gff-density}
\end{align}

To characterize the low-$t$ behavior, we introduce the concept of the mean square radii defined through the slopes of the GFFs at $t=0$,
\begin{align}
  \langle r^2\rangle_F&=\frac{6}{F(0)}
    \left.\frac{\dd F(t)}{\dd t}\right|_{t=0},
    \qquad F=A,D,\theta.
    \label{eq:gff-radii}
\end{align}

\section{General construction of a conserved EMT in the Skyrme model}\label{sec:skyrme}
\subsection{Skyrme model}
\label{sec:basic-skyrme}

In this subsection, we summarize the Skyrme model and its large-$\Nc$ expansion as the basis for the order-by-order construction of a conserved EMT in Sec.~\ref{sec:construction}.

We start with the conventional Skyrme Lagrangian with a pion mass term:
\begin{align}
  \mathcal L_{\pi}={}&\frac{f_\pi^2}{4}
    \mathrm{tr}(\partial_\mu U_\pi^\dagger\partial^\mu U_\pi) +\frac{1}{32e^2}\mathrm{tr} ([ U_{\pi}^{\dagger} \partial_{\mu} U_{\pi} ,  U_{\pi}^{\dagger} \partial_{\nu} U_{\pi} ]^{2}) \notag\\
  &+\frac{f_\pi^2m_\pi^2}{4}
    \mathrm{tr}(U_\pi+U_\pi^\dagger-2),
  \label{eq:skyrme-pion-lagrangian}
\end{align}
where $f_\pi$ is the pion decay constant, $m_\pi$ is the pion mass, and $e$ is the Skyrme parameter.
The SU(2) chiral field $U_\pi$ is parametrized as
\begin{align}
    U_\pi(x)=\exp[i\bm\tau\cdot\bm\pi(x)/f_\pi] \ , \label{eq:pion-matrix}
\end{align}
where $\bm\tau$ are the Pauli matrices and $\bm\pi(x)$ denote the pion fields.
The trace is taken over flavor indices.
To obtain the soliton solution, we adopt the hedgehog ansatz in the rest frame:
\begin{equation}
  U_\pi(\bm r)=\exp[i\bm\tau\cdot\hat{\bm r}F(r)]
  \ ,
  \label{eq:hedgehog-background}
\end{equation}
where we  define  $\hat{\bm r}={\bm r}/r$.
$F(r)$ is the profile function satisfying the following boundary conditions
\begin{equation}
F(0)=\pi,\, F(\infty)=0 \ ,
\end{equation}
which ensure a unit baryon number.
The hedgehog configuration is invariant under combined spatial and isospin rotations.
This combined rotational symmetry is referred to as ``hedgehog symmetry''.
We substitute the hedgehog ansatz~\eqref{eq:hedgehog-background} into the Lagrangian~\eqref{eq:skyrme-pion-lagrangian} to define the action as $S_{\pi}$.
Then, varying the action with respect to the profile function $F$ as
\begin{equation}
  \frac{\delta S_{\pi}}{\delta F} = 0,
  \label{eq:leading-skyrme-eom}
\end{equation}
we obtain the EOM for $F$ as
\begin{align}
 &\left(r^2+\frac{2\sin^2F}{e^2f_\pi^2}\right)F''
   +2rF'\notag\\
 & \quad + \sin2F \left[
   \frac{1}{e^2f_\pi^2}\left(F'^2-\frac{\sin^2F}{r^2}\right)-1
   \right]\notag\\
 &\quad-m_\pi^2r^2\sin F=0,
 \label{eq:leading-skyrme-eom-explicit}
\end{align}
where primes denote derivatives with respect to $r$.
By substituting the solution of the above EOM into the Lagrangian, the soliton mass is calculated as
\begin{equation}
M_{\text{Skyrme}}[F] = - 4\pi\int\dd r\, r^2 \mathcal{L}_\pi[F] \ .
\end{equation}

\begin{widetext}
From the Lagrangian in Eq.~\eqref{eq:skyrme-pion-lagrangian}, the
EMT in the Skyrme model is given by
\begin{align}
  T_{\pi, \mu\nu} = &\frac{f_\pi^2}{2} \mathrm{tr} \left( \partial_{\mu} U_{\pi}^{\dagger} \partial_{\nu} U_{\pi} \right)
  +\frac{1}{8e^2} \mathrm{tr} \left( [ U_{\pi}^{\dagger} \partial_{\mu} U_{\pi} ,  U_{\pi}^{\dagger} \partial_{\alpha} U_{\pi} ] [ U_{\pi}^{\dagger} \partial_{\nu} U_{\pi} ,  U_{\pi}^{\dagger} \partial^{\alpha} U_{\pi} ] \right)\notag\\
  &-g_{\mu\nu}\left\{\frac{f_\pi^2}{4} \mathrm{tr} \left(\partial_\alpha U_\pi^\dagger\partial^\alpha U_\pi \right)
    +\frac{1}{32e^2}\mathrm{tr} \left( [ U_{\pi}^{\dagger} \partial_{\alpha} U_{\pi} ,  U_{\pi}^{\dagger} \partial_{\beta} U_{\pi} ]^{2} \right)
    +\frac{f_\pi^2m_\pi^2}{4}
      \mathrm{tr}(U_\pi+U_\pi^\dagger-2)\right\}.
    \label{eq:pion-covariant-emt}
\end{align}
We note that the parameters scale with $\Nc$ as~\cite{Cebulla:2007ei} 
\begin{equation}
  f_\pi=\order{\Nc^{1/2}},\quad
  e=\order{\Nc^{-1/2}},\quad
  m_\pi=\order{\Nc^0}.
  \label{eq:skyrme-counting}
\end{equation}
Thus $S_\pi$, $M_\text{Skyrme}$, and $T_\pi^{\mu\nu}$ are of $\order{\Nc}$.
Substituting the hedgehog ansatz into the above expression, we can easily show that the resultant EMT satisfies the conservation law. 
From Eqs.~\eqref{eq:energy-density} and \eqref{eq:spatial-emt-densities}, we obtain the following equations:
\begin{align}
  \Eden_\pi[F]
    &=\frac{f_\pi^2}{2}\left(F'^2+\frac{2\sin^2F(r)}{r^2}\right)
      +\frac{\sin^2F(r)}{2e^2r^2}\left(2F'^2+\frac{\sin^2F(r)}{r^2}\right)
      +f_\pi^2m_\pi^2\bigl(1-\cos F(r)\bigr),
      \label{eq:model-pion-energy}\\
  p_\pi[F]
    &=-\frac{f_\pi^2}{6}\left(F'^2+\frac{2\sin^2F(r)}{r^2}\right)
      +\frac{\sin^2F(r)}{6e^2r^2}\left(2F'^2+\frac{\sin^2F(r)}{r^2}\right)
      -f_\pi^2m_\pi^2\bigl(1-\cos F(r)\bigr),
      \label{eq:model-pion-trace}\\
  s_\pi[F]
    &=\left(f_\pi^2+\frac{\sin^2F(r)}{e^2r^2}\right)
      \left(F'^2-\frac{\sin^2F(r)}{r^2}\right).
      \label{eq:model-pion-traceless}
\end{align}

\end{widetext}

\subsection{Conserved EMT in the large-$N_c$ expansion with additional dynamical fields}
\label{sec:construction}

In this subsection, we discuss the general construction of a conserved EMT in the bound-Skyrmion model where the Skyrmion is coupled to additional dynamical fields.
We first show that the fixed-background approximation violates the conservation of the EMT and is inconsistent with the large-$\Nc$ expansion.
We then construct a conserved EMT that resolves these issues.

We first introduce $\varphi^a = \pi^{a} / f_{\pi}$ (with $a=1,2,3$) to parameterize $U = \mathrm{exp}( i \bm{\tau} \cdot \bm{\varphi} ) $, so that a variation can include nonspherical components rather than only a radial profile.
Then, we consider the following action in which the additional dynamical fields $\psi = \{ \psi^{1}, \psi^{2}, \cdots \}$ are coupled to the Skyrme model:
\begin{equation}
  S[\varphi,\psi]=S_{\pi}[\varphi]+S_I[\varphi,\psi] .
  \label{eq:general-coupled-action}
\end{equation}
The term $S_I$ includes the kinetic, mass, and interaction terms for $\psi$.
We assume $S_{I}$ is the sub-leading term relative to $S_{\pi}$ in the $1/\Nc$ expansion as:
\begin{align}
  S_{\pi}[\varphi] = \order{\Nc}, \quad
  S_I[\varphi,\psi] = \order{\Nc^0}.
  \label{eq:general-coupled-action-order}
\end{align}

The equations of motion (EOMs) for $\varphi$ and $\psi$ are obtained from $S[\varphi, \psi]$ in Eq.~(\ref{eq:general-coupled-action}) as
\begin{align}
& \frac{\delta S[\varphi,\psi]}{\delta \varphi^a}= \frac{\delta S_\pi[\varphi]}{\delta \varphi^a} + \frac{\delta S_I[\varphi,\psi]}{\delta \varphi^a} = 0 \ , 
\label{EOM: varphi 1} \\
& \frac{\delta S[\varphi,\psi]}{\delta \psi^{i}} = \frac{\delta S_I[\varphi,\psi]}{\delta \psi^{i}}  = 0 \ .
\label{EOM: psi 1}
\end{align}
From the large-$N_c$ scaling in Eq.~(\ref{eq:general-coupled-action-order}), the second term in eq.~\eqref{EOM: varphi 1} is $1/N_c$ suppressed compared with the first term.
Then, in the fixed-background approximation, we first solve the EOM in \eqref{EOM: varphi 1} without the second term $\frac{\delta S_I[\varphi,\psi]}{\delta \varphi^a}$, and obtain $\varphi_{0}= \{\varphi^1_0, \varphi^2_0, \varphi^3_0 \}$ as a solution. After that, fixing $\varphi=\varphi_0$, we solve the EOM~\eqref{EOM: psi 1}.
In summary, the following two EOMs for $\varphi$ and $\psi$ are solved in the fixed-background approximation,
\begin{equation}
  \left.\frac{\delta S_{\pi}[\varphi]}{\delta \varphi^{a}}\right|_{\varphi_0}=0, \label{eq:eom-pion-fixed}
\end{equation}
and
\begin{equation}
  \left.\frac{\delta S_I[\varphi=\varphi_0,\psi]}{\delta \psi^{i}}\right|_{\psi_0}=0,
  \label{eq:eom-additional-fixed}
\end{equation}
where $\varphi_{0}$ and $\psi_{0} = \{\psi^1_0, \psi^2_0, \cdots \}$ are the solutions of Eqs.~\eqref{eq:eom-pion-fixed} and \eqref{eq:eom-additional-fixed}, respectively.
We note that the second term in Eq.~(\ref{EOM: varphi 1}) generates the backreaction from the additional fields to the pion field.
Therefore, the fixed-background approximation implies that the backreaction is neglected.

We would like to stress that the fixed-background approximation properly provides the mass of the bound state up to ${\mathcal O}(N_c^0)$ (next-to-leading order).
However, this approximation neglecting the backreaction violates the conservation law of the EMT.
To explaining this, we note that the divergence of the EMT obeys the following identity:
\begin{align}
  \partial_\mu T^{\mu\nu} =& -\sum_A\frac{\delta S}{\delta \phi^A(x)}\partial^\nu \phi^A \notag \\
  =& - \sum_{a} \frac{ \delta S_{\pi} }{ \delta \varphi^a } \partial^\nu \varphi^a
     - \sum_{a} \frac{ \delta S_{I} }{ \delta \varphi^a } \partial^\nu \varphi^a
     - \sum_{i} \frac{ \delta S_{I} }{ \delta \psi^{i} } \partial^\nu \psi^{i},
  \label{eq:emt-eom-identity}
\end{align}
where $\phi^A$ runs over all fields, including $\varphi$ and $\psi$.
Since Eqs.~\eqref{eq:eom-pion-fixed} and \eqref{eq:eom-additional-fixed} only guarantee the vanishing of the first and last terms in Eq.~\eqref{eq:emt-eom-identity} but not the second term corresponding to the backreaction, the EMT is not conserved in the fixed-background approximation.
Thus, the effects of backreaction to the EMT cannot be neglected.
This observation motivates us to construct a conserved EMT by consistently incorporating the backreaction to the pion field.

To preserve the conserved EMT, each term of the EMT should be treated consistently with the large-$\Nc$ expansion.
We expand the pion field $\varphi^{a}$ and the additional fields $\psi^{i}$ in powers of $1/\Nc$ as
\begin{align} 
  \varphi^{a}&=\sum_{n=0}^\infty \varphi_n^{a} ,\quad \varphi^{a}_{n}=\order{\Nc^{-n}} ,
  \label{eq:general-field-expansion}
\\
  \psi^{i}&=\sum_{n=0}^\infty \psi_n^{i} ,\quad \psi^{i}_{n}=\order{\Nc^{-n}}.
  \label{eq:general-field-expansion-psi}
\end{align}
Substituting Eqs.~\eqref{eq:general-field-expansion} and \eqref{eq:general-field-expansion-psi} into the full EOMs in Eqs.~(\ref{EOM: varphi 1}) and (\ref{EOM: psi 1}), and expanding them in terms of $1/N_c$, we obtain
\begin{align}
  0={}&\left.\frac{\delta S_\pi[\varphi]}{\delta\varphi^a(x)}\right|_{\varphi_0}
  \notag\\
  &+(\mathcal D\varphi_1)^a(x)+J^a(x)
  +\order{\Nc^{-1}},
  \label{eq:general-expanded-pion-eom}
\end{align}
and
\begin{align}
  0={}&\left.\frac{\delta S_I[\varphi=\varphi_0,\psi]}{\delta\psi^i(x)}\right|_{\psi_0}
  +\order{\Nc^{-1}},
  \label{eq:general-expanded-psi-eom}
\end{align}
where $\mathcal D$ is the linearized operator associated with the leading Skyrme action and $J^a$ is the source generated by the additional sector, which are defined as
\begin{align}
 (\mathcal D \varphi_1)^a(x)
 &\equiv\int\dd^4y\,
 \left.\frac{\delta^2S_{\pi}[\varphi]}
 {\delta\varphi^a(x)\delta\varphi^b(y)}\right|_{\varphi_0}
 \varphi_1^b(y),\notag\\
 J^a(x)&\equiv\left.\frac{\delta S_I[\varphi=\varphi_0,\psi]}{\delta\varphi^a(x)}
 \right|_{\psi_0}.
 \label{eq:general-backreaction}
\end{align}
Since the terms of order $\Nc$ and $\Nc^0$ must vanish separately, we obtain
\begin{align}
  \order{\Nc}:&\quad
  \left.\frac{\delta S_\pi[\varphi]}{\delta\varphi^a(x)}\right|_{\varphi_0}=0,
  \label{eq:general-pion-eom-lo}\\
  \order{\Nc^0}:&\quad
  (\mathcal D\varphi_1)^a(x)=-J^a(x),
  \label{eq:general-pion-eom-nlo}
\end{align}
and
\begin{align}
  \order{\Nc^0}:&\quad
  \left.\frac{\delta S_I[\varphi=\varphi_0,\psi]}{\delta\psi^i(x)}\right|_{\psi_0}=0.
  \label{eq:general-psi-eom-nlo}
\end{align}

Let us now consider the EMT in the large-$\Nc$ expansion,
\begin{align}
  T^{\mu\nu}
    &=T_0^{\mu\nu}+T_1^{\mu\nu}+\order{\Nc^{-1}},
\end{align}
where
\begin{align}
  T_0^{\mu\nu}&=\order{\Nc},\qquad
  T_1^{\mu\nu}=\order{\Nc^0}.
  \label{eq:general-emt-expansion}
\end{align}
Here $T_0^{\mu\nu}$ is obtained by substituting $U_\pi = \exp\left( i\bm{\tau}\cdot\bm{\varphi}_0\right)$ into the EMT of the Skyrme model given in Eq.~\eqref{eq:pion-covariant-emt}:
\begin{equation}
T_0^{\mu\nu} = T_\pi^{\mu\nu}[\varphi=\varphi_0] .
\end{equation}
The next-to-leading contribution $T_1^{\mu\nu}$ is expressed as
\begin{align}
  T_1^{\mu\nu}(x)
    &=\int\dd^4y\,
    \left.\frac{\delta T_{\pi}^{\mu\nu}[\varphi](x)}
    {\delta\varphi^a(y)}\right|_{\varphi_0}\varphi_1^a(y)
      +T_I^{\mu\nu}[\varphi_0,\psi_0](x),
  \label{eq:general-first-emt-raw}
\end{align}
where $T_{I}^{\mu\nu}$ is the EMT obtained from $S_{I}$.
We note that, although both terms in Eq.~(\ref{eq:general-first-emt-raw}) are of ${\mathcal O}(N_c^0)$, only the second term is kept in the fixed-background approximation.
As a result, the EMT does not satisfy the conservation law in the fixed-background approximation.
Actually, by using the EOMs~\eqref{eq:general-pion-eom-lo}-\eqref{eq:general-psi-eom-nlo} and the identity~\eqref{eq:emt-eom-identity}, the divergences of the terms in Eq.~(\ref{eq:general-first-emt-raw}) are calculated as
\begin{align}
 \partial_\mu T_I^{\mu\nu}
 &=-J_a\partial^\nu\varphi_0^a,
\end{align}
and
\begin{align}
 \partial_\mu 
 \left.\frac{\delta T_{\pi}^{\mu\nu}[\varphi]}
    {\delta\varphi^a}\right|_{\varphi_0}\varphi_1^a
 &=-(\mathcal D\varphi_1)_a\partial^\nu\varphi_0^a\notag\\
 &=J_a\partial^\nu\varphi_0^a.
 \label{eq:general-first-cancellation}
\end{align}
This implies that the two divergences exactly cancel with each other, and that the EMT $T_1^{\mu\nu}$ is conserved.

Before proceeding, we emphasize that the fixed-background approximation remains sufficient for the mass through $\order{\Nc^0}$.
Since the soliton is static,  we can show that 
the contribution to the energy from the backreaction to the pion field, which is proportional to $\varphi_1$, vanishes when integrated over the space:
\begin{equation}
\int d^3 x 
\left.\frac{\delta T_{\pi}^{00}[\varphi]}
    {\delta\varphi^a}\right|_{\varphi_0}\varphi_1^a = 0 .
\label{eq:general-energy-variation-zero}
\end{equation}
The total mass therefore takes the form
\begin{equation}
  M=M_0[\varphi_0]
    +\int\dd^3x\,T_I^{00}[\varphi_0,\psi_0](\bm x)
    +\order{\Nc^{-1}}.
  \label{eq:general-mass-through-nlo}
\end{equation}

Since the rotational energy is of $\order{\Nc^{-1}}$~\cite{Cebulla:2007ei}, the backreaction changes the rotational energy by $\order{\Nc^{-2}}$.
Hence, the mass splittings among the $\Lambda$, $\Sigma$, and $\Sigma^\ast$ derived in Ref.~\cite{Callan:1985hy} are not changed by the backreaction through $\order{\Nc^{-1}}$.

The above construction can be extended to higher orders by expanding EOMs and the EMTs consistently in the large-$N_c$ expansion.
At each order, the correction to each field obeys a linear equation with the source built from lower-order solutions.
Corrections to the eigenenergy must also be included at the required order.
Using Eq.~\eqref{eq:emt-eom-identity}, we can verify that the EMT is conserved order by order.
The method presented in this subsection allows us to specify explicitly the order in $1/\Nc$ up to which contributions to the EMT are retained.

\section{Application to hyperons in the bound-state approach}
\label{sec:model}

We now apply the construction of the EMT given in Sec.~\ref{sec:construction} to the ``Callan-Klevanov (CK) bound-state approach''~\cite{Callan:1985hy,Callan:1987xt} in which a hyperon is described as the bound-state of an antikaon and an $\mathrm{SU}(2)$ Skyrmion.
Collective rotation is not included in this work.

\subsection{Bound-state approach}

In this subsection, we briefly review the CK bound-state approach.
The action~\footnote{In Refs~\cite{Callan:1985hy,Callan:1987xt}, the pion decay constant is  denoted by $F_\pi$ which is related to $f_\pi$ by $F_\pi = 2 f_\pi$ in the convention used here.} is~\cite{Callan:1985hy,Callan:1987xt}
\begin{align}
  S =& \int \dd^{4} x \left( \mathcal{L}_{SU(3)} + \mathcal{L}_{\text{SB}} \right) + S_{\rm WZ}, 
\end{align}
where
\begin{align}
  \mathcal{L}_{SU(3)} =& \frac{f_{\pi}^{2}}{4} \text{Tr}(\partial_{\mu} U \partial^{\mu} U^{\dagger}) \notag \\
  &+ \frac{1}{32 e^{2}} \text{Tr}([U^{\dagger}\partial_{\mu} U, U^{\dagger}\partial_{\nu} U]^{2}) , \notag \\
  \mathcal{L}_{\text{SB}} =& \frac{1}{12} f_{\pi}^{2} \left( m_{\pi}^{2} + 2 m_{K}^{2} \right) \text{Tr}(U + U^{\dagger} - 2) \notag \\
  & + \frac{\sqrt{3}}{6} f_{\pi}^{2} \left( m_{\pi}^{2} - m_{K}^{2} \right) \text{Tr}(\lambda_{8}(U + U^{\dagger})) , \label{eq:skyrme-action}
\end{align}
and
$S_{\rm WZ}$ is the Wess-Zumino action, $U$ is the $\mathrm{SU}(3)$ chiral field, and $m_K$ is the kaon mass.
The Gell-Mann matrix $\lambda_8$ is
\begin{align}
  \lambda_{8} = \frac{1}{\sqrt{3}} \begin{pmatrix}
    1 & 0 & 0 \\
    0 & 1 & 0 \\
    0 & 0 & -2
  \end{pmatrix}.
\end{align}
Now, we employ the Callan-Klebanov ansatz~\cite{Callan:1985hy,Callan:1987xt} for the $\mathrm{SU(3)}$ chiral field, which is given by
\begin{equation}
  U = \Xi U_{K} \Xi ,
\end{equation}
where
\begin{align}
    \Xi = \mathrm{diag}(\xi, 1), \; \xi^2 = U_\pi .
\end{align}
Here $U_\pi$ is the SU(2) matrix given in Eq.~\eqref{eq:pion-matrix}, and $U_{K}$ is the kaon field, which is given by
\begin{align}
  U_{K} = \exp\left(\frac{i \sqrt{2}}{f_{\pi}} \begin{pmatrix}
    0 & K \\
    K^{\dagger} & 0
  \end{pmatrix}\right) .
\end{align}
The kaon iso-doublet field is
\begin{equation}
K = \begin{pmatrix}
K^+ \\ K^0 
\end{pmatrix} .
\end{equation}

\begin{widetext}
In the CK bound-state approach, we treat the kaon field as a small fluctuation around the pion background, and we expand the action up to second order in the kaon field as
\begin{align}
  \mathcal{L}_{SU(3)} &= \mathcal{L}_{SU(2)} + \mathcal{L}_{K} + \mathcal{O}(K^{4}), \notag \\
  \mathcal{L}_{\text{SB}}  &
  = \frac{f_{\pi}^{2} m_{\pi}^{2}}{4} \text{Tr}(U_{\pi} + U_{\pi}^{\dagger} - 2)
  - m_{K}^{2} K^{\dagger} K
  + \frac{m_{\pi}^{2}}{4} K^{\dagger} \left( 2 - U_{\pi}^{\dagger} - U_{\pi} \right) K,
  \label{eq:sb-lagrangian}
\end{align}
where
\begin{align}
  \mathcal{L}_{SU(2)} &= \frac{f_{\pi}^{2}}{4} \text{Tr}(\partial_{\mu} U_{\pi} \partial^{\mu} U_{\pi}^{\dagger})
  + \frac{1}{32 e^{2}} \text{Tr}([U_{\pi}^{\dagger}\partial_{\mu} U_{\pi}, U_{\pi}^{\dagger}\partial_{\nu} U_{\pi}]^{2}), \label{eq:pion-lagrangian} \\
  \mathcal{L}_{K} &= (D_{\mu} K)^{\dagger} D^{\mu} K - K^{\dagger} a_{\mu}^{\dagger} a^{\mu} K \notag \\
  &\quad + \frac{1}{(2 e f_{\pi})^{2}} \Bigl\{
  - \frac{1}{8}K^{\dagger}K \mathrm{tr}[\partial_{\mu}U_{\pi} U_{\pi}^{\dagger}, \partial_{\nu}U_{\pi} U_{\pi}^{\dagger}]^{2}
  - 2 (D_{\mu}K)^{\dagger} D_{\nu} K \mathrm{tr} (a^{\mu} a^{\nu}) \notag \\
  &\qquad - \frac{1}{2} (D_{\mu}K)^{\dagger} D^{\mu}K \mathrm{tr}(\partial_{\nu}U_{\pi}^{\dagger} \partial^{\nu}U_{\pi})
  + 6 (D_{\nu}K)^{\dagger} [a^{\nu}, a^{\mu}] D_{\mu}K \Bigr\},
  \label{eq:kaon-lagrangian} 
\end{align}
\end{widetext}
with 
\begin{align}  
  a_\mu&=\tfrac12(\xi^\dagger\partial_\mu\xi
  -\xi\partial_\mu\xi^\dagger).
\end{align}
In the above expression, the
covariant derivative for the $K$ field is defined by
\begin{align}
  D_\mu K&=(\partial_\mu+v_\mu)K,
\end{align}
where
\begin{align}
  v_\mu&=\tfrac12(\xi^\dagger\partial_\mu\xi
    +\xi\partial_\mu\xi^\dagger) .
  \label{eq:emt-kaon-connection}
\end{align}
We also obtain the Wess-Zumino term up to second order in the kaon field as
\begin{align}
  \mathcal{L}_{\text{WZ}} =& - i \frac{N_{c}}{4 f_{\pi}^{2}} B^{\mu} \left[ \left( D_{\mu} K \right)^{\dagger} K - K^{\dagger} D_{\mu} K \right],
  \label{eq:wz-lagrangian} 
\end{align}
where the baryon number current $B^\mu$ is given by
\begin{align}
  B^{\mu} =& - \frac{\epsilon^{\mu\nu\alpha\beta}}{24 \pi^{2}} \text{Tr} \left[ (U_{\pi}^{\dagger} \partial_{\nu} U_{\pi}) (U_{\pi}^{\dagger} \partial_{\alpha} U_{\pi}) (U_{\pi}^{\dagger} \partial_{\beta} U_{\pi}) \right]. 
\end{align}
In this work, we will use the action for kaons truncated at quadratic order in the kaon field:
\begin{align}
S_K = & \int d^4 x \bigg[
{\mathcal L}_K + {\mathcal L}_{\rm WZ}
\notag\\
& \qquad \quad
 - m_{K}^{2} K^{\dagger} K
  + \frac{m_{\pi}^{2}}{4} K^{\dagger} \left( 2 - U_{\pi}^{\dagger} - U_{\pi} \right) K
\bigg] . 
\end{align}
In this application, the action for the additional sector denoted as $S_I$ in Sec.~\ref{sec:construction} is identified as 
\begin{align}
  S_I=S_K.
\end{align}
They scale with $N_{c}$ as 
\begin{align}
  m_{K} =& \order{\Nc^0},\quad K =& \order{\Nc^0},\quad
  S_{K} =& \order{\Nc^0}. \label{eq:kaon-counting}
\end{align}

Because of the hedgehog symmetry, the kaon modes can be classified by the grand spin $\bm T=\bm L+\bm I$, where $\bm L$ and $\bm I$ are the orbital angular momentum and the isospin of the kaon, respectively, with the isospin quantum number $I=1/2$~\cite{Callan:1985hy,Callan:1987xt}.
We consider one antikaon with  the strangeness $S=-1$ in the ground-state $L=1$ and $T=1/2$.
We separate the variables of the kaon field as follows:
\begin{align}
  K(t,\bm r)=e^{-iE_Kt}k(r)Y_{TLT_z}(\hat{\bm r}),
  \label{eq:kaon-mode}
\end{align}
where the angular wave function $Y_{TLT_z}$ is a two-component vector in isospin space, obtained by coupling the orbital angular momentum $L$ and isospin $I=1/2$ to grand spin $T$ with $T_z$.
We normalize it by $\int\dd\Omega\,Y_{TLT_z}^\dagger Y_{TLT_z}=1$.

Substituting Eq.~\eqref{eq:kaon-mode} into $S_K$ and performing the angular integration, we obtain the equation of motion for the radial wavefunction as
\begin{align}
  &-\frac{1}{r^2}\frac{\dd}{\dd r}
    \left(r^2h(r)\frac{\dd k(r) }{\dd r}\right)\notag\\
  &\quad+\Bigl[m_K^2-m_\pi^2\sin^2\frac{F}{2}
    +V_{\mathrm{eff}}(r)\notag\\
  &\qquad\qquad-E_K^2f(r)-2E_K\lambda(r)\Bigr]k(r)=0,
  \label{eq:kaon-radial-eom}
\end{align}
where $E_K$ is the energy eigenvalue,
\begin{align}
  f(r)&=1+\frac{1}{(2ef_\pi)^2}
    \left(F'^2+\frac{2\sin^2F}{r^2}\right),\notag\\
  \lambda(r)&=-\frac{\Nc}{8\pi^2f_\pi^2}
    \frac{\sin^2F}{r^2}F',\notag\\
  h(r)&=1+\frac{2}{(2ef_\pi)^2}
    \frac{\sin^2F}{r^2},
  \label{eq:kaon-radial-coefficients}
\end{align}
and the effective potential $V_{\rm eff}(r)$ for the channel with $L=1$ and $T=1/2$ is given by
\begin{align}
 V_{\mathrm{eff}}(r)
 ={}&-\frac14\left(F'^2+\frac{2\sin^2F}{r^2}\right)
       +\frac{2\cos^4(F/2)}{r^2}\notag\\
 &+\frac{1}{(2ef_\pi)^2}\Biggl[
    -\frac{2\sin^2F}{r^2}
       \left(2F'^2+\frac{\sin^2F}{r^2}\right)\notag\\
 &\qquad+\frac{2\cos^4(F/2)}{r^2}
       \left(F'^2+\frac{4\sin^2F}{r^2}\right)\notag\\
 &\qquad-\frac{6}{r^2}\frac{\dd}{\dd r}
       \left(F' \cos^2\frac F2\,\sin F\right)\Biggr].
 \label{eq:kaon-effective-potential}
\end{align}
The normalization condition for the radial
wavefunction
is~\cite{Callan:1987xt}
\begin{align}
  2\int_0^\infty\dd r\,r^2
  [E_Kf(r)+\lambda(r)]|k(r)|^2=1.
  \label{eq:kaon-normalization}
\end{align}

\subsection{EOM in large-$\Nc$ expansion}\label{sec:backreaction}

Let us apply the method described in Sec.~\ref{sec:construction} to the bound system of an antikaon and a Skyrmion.
To include the backreaction, we expand the profile function for pion and the radial wave function for kaon in powers of $1/\Nc$:
\begin{gather}
  F(r) = \sum_{n=0} F_n(r) , \quad F_{n}(r)=\order{\Nc^{-n}}, \label{eq:profile-backreaction}
\\
  k(r) = \sum_{n=0} k_n(r) , \quad k_{n}(r)=\order{\Nc^{-n}}. \label{eq:kaon-backreaction}
\end{gather}
The kaon eigenenergy is expanded consistently as
\begin{gather}
  E_K = \sum_{n=0} E_{K,n}, \quad E_{K,n}=\order{\Nc^{-n}}.
  \label{eq:kaon-energy-expansion}
\end{gather}
For fixed $T$, $L$, and $T_z$, the angular wavefunction $Y_{TLT_z}$ is independent of $\Nc$ and the backreaction considered here preserves the hedgehog symmetry.
Therefore, the angular wavefunction $Y_{TLT_z}$ remains unchanged here.

The full EOMs are obtained as
\begin{align}
  \frac{\delta S}{\delta F(r)}
  & =\frac{\delta S_\pi}{\delta F(r)}
   +\frac{\delta S_K}{\delta F(r)}=0,
  \label{eq:full-profile-eom}
\\
  \frac{\delta S}{\delta k(r)}
  &=\frac{\delta S_K}{\delta k(r)}=0,
  \label{eq:full-kaon-eom}
\end{align}
where the functional derivatives are taken after angular integration.
Substituting Eqs.~\eqref{eq:profile-backreaction} and \eqref{eq:kaon-backreaction} into the above EOMs, and expanding through $\order{\Nc^0}$, we obtain
\begin{align}
  0={}&\left.\frac{\delta S_\pi}{\delta F(r)}\right|_{F_0}
  \notag\\
  &+(\mathcal D_\pi F_1)(r)+J_K(r)
  +\order{\Nc^{-1}},
  \label{eq:expanded-profile-eom}
\end{align}
and
\begin{align}
  0={}&\left.\frac{\delta S_K}{\delta k(r)}\right|_{F_0,k_0} + \order{\Nc^{-1}} ,
  \label{eq:expanded-kaon-eom}
\end{align}
where the linearized operator and the kaon source are defined by
\begin{align}
  (\mathcal D_\pi F_1)(r)
  &\equiv\int_0^\infty\dd r'\,
  \left.\frac{\delta^2S_\pi}
  {\delta F(r)\delta F(r')}\right|_{F_0}F_1(r'),
  \label{eq:linearized-pion-operator}\\
  J_K(r)
  &\equiv\left.\frac{\delta S_K}{\delta F(r)}
  \right|_{F_0,k_0}.
  \label{eq:kaon-profile-source}
\end{align}
The functional derivative defining $J_K$ is taken at fixed $k_0$ and $E_{K,0}$.
Since the terms of order $\Nc$ and $\Nc^0$ must vanish separately, we obtain
\begin{align}
  \order{\Nc}:&\quad
  \left.\frac{\delta S_\pi}{\delta F(r)}\right|_{F_0}=0,
  \label{eq:profile-eom-lo}\\
  \order{\Nc^0}:&\quad
  (\mathcal D_\pi F_1)(r)=-J_K(r),
  \label{eq:backreaction-eom}
\end{align}
and
\begin{align}
  \order{\Nc^0}:&\quad
  \left.\frac{\delta S_K}{\delta k(r)}\right|_{F_0,k_0}=0.
  \label{eq:kaon-eom-lo}
\end{align}

\subsection{EMT in large-$\Nc$ expansion}\label{sec:bound-state-emt}

The total EMT is expanded as
\begin{equation}
T^{\mu\nu} = \sum_{n=0} T_{n}^{\mu\nu} ,
\label{eq:bound-state-emt-nlo}
\end{equation}
where
$T_n^{\mu\nu}$ is of ${\mathcal O}(N_c^{1-n})$.
The leading-order contribution is obtained from the pion field as
\begin{equation}
T_{0}^{\mu\nu} = T_\pi^{\mu\nu}[F_0],
\end{equation}
where
$T_\pi^{\mu\nu}[F_0]$ is given in 
Eq.~(\ref{eq:pion-covariant-emt}).
The contribution at the next-to-leading order is expressed as
\begin{equation}
T_{1}^{\mu\nu} = T_{\pi,1}^{\mu\nu}[F_0,F_1] + T_K^{\mu\nu}[F_0,k_0],
\end{equation}
where
\begin{align}
  T_{\pi,1}^{\mu\nu}[F_0,F_1] & = \int_0^\infty\dd r'\,
    \left.\frac{\delta T_\pi^{\mu\nu}[F](\bm r)}
    {\delta F(r')}\right|_{F_0}F_1(r') ,
  \label{eq:linearized-pion-emt}
\\
T_{K, \mu\nu}[F_0,k_0] & = 
T^{(2)}_{K, \mu\nu}+T^{(4)}_{K, \mu\nu}
\nonumber\\
& \ {}+ g_{\mu\nu}\Bigl[m_K^2K^\dagger K
    -\frac{m_\pi^2}{4}K^\dagger
      (2-U_\pi-U_\pi^\dagger)K\Bigr] ,
\end{align}
with
\begin{widetext}
\begin{align}
  T^{(2)}_{K, \mu\nu}
  ={}&(D_\mu K)^\dagger D_\nu K+(D_\nu K)^\dagger D_\mu K
    -K^\dagger(a_\mu^\dagger a_\nu+a_\nu^\dagger a_\mu)K - g_{\mu\nu}\left[(D_\alpha K)^\dagger D^\alpha K
    -K^\dagger a_\alpha^\dagger a^\alpha K\right],
    \label{eq:kaon-two-derivative-emt}\\
  T^{(4)}_{K, \mu\nu}
  ={}&\frac{1}{(2ef_\pi)^2}\Bigl\{
    -\frac12 K^\dagger K\,\mathrm{tr}
      ([U_\pi^\dagger\partial_\mu U_\pi,U_\pi^\dagger\partial_\alpha U_\pi][U_\pi^\dagger\partial_\nu U_\pi,U_\pi^\dagger\partial^\alpha U_\pi])
    \notag\\
  &\quad-2\bigl[(D_\mu K)^\dagger D_\alpha K\,
       \mathrm{tr}(a_\nu a^\alpha)
    +(D_\nu K)^\dagger D_\alpha K\,
       \mathrm{tr}(a_\mu a^\alpha) + (D_\alpha K)^\dagger D_\mu K\,
       \mathrm{tr}(a^\alpha a_\nu)
    +(D_\alpha K)^\dagger D_\nu K\,
       \mathrm{tr}(a^\alpha a_\mu)\bigr] \notag \\
  &\quad-\frac12\bigl[
    \bigl((D_\mu K)^\dagger D_\nu K
         +(D_\nu K)^\dagger D_\mu K\bigr)\mathrm{tr}(\partial_\alpha U_\pi^\dagger\partial^\alpha U_\pi) + 2(D_\alpha K)^\dagger D^\alpha K\,\mathrm{tr}(\partial_\mu U_\pi^\dagger\partial_\nu U_\pi)\bigr]
    \notag\\
  &\quad+6\bigl[
     (D_\alpha K)^\dagger[a^\alpha,a_\mu]D_\nu K
    +(D_\alpha K)^\dagger[a^\alpha,a_\nu]D_\mu K + (D_\mu K)^\dagger[a_\nu,a^\alpha]D_\alpha K
    +(D_\nu K)^\dagger[a_\mu,a^\alpha]D_\alpha K\bigr] \notag \\
  &\quad+g_{\mu\nu}\Bigl[
    \frac18 K^\dagger K\,\mathrm{tr}
      ([U_\pi^\dagger\partial_\alpha U_\pi,U_\pi^\dagger\partial_\beta U_\pi][U_\pi^\dagger\partial^\alpha U_\pi,U_\pi^\dagger\partial^\beta U_\pi]) + 2(D_\alpha K)^\dagger D_\beta K\, \mathrm{tr}(a^\alpha a^\beta) \notag \\
  &\qquad \qquad  +\frac12(D_\alpha K)^\dagger D^\alpha K\,
       \mathrm{tr}(\partial_\beta U_\pi^\dagger\partial^\beta U_\pi) - 6(D_\alpha K)^\dagger[a^\alpha,a^\beta]
       D_\beta K\Bigr]\Bigr\}.
    \label{eq:kaon-four-derivative-emt}
\end{align}
\end{widetext}
We would like to stress that, using the EOMs~\eqref{eq:backreaction-eom} and \eqref{eq:kaon-eom-lo}, one can easily show that $T_0^{\mu\nu}$ and $T_1^{\mu\nu}$ are separately conserved.

Let us comment on the angular dependence of the contribution to the EMT from the kaon field.
This is expected to be spherically symmetric for $L=1$ and $T=1/2$, even though the kaon wavefunction has a nontrivial angular dependence.
This can be seen as follows:
The angular wavefunction for $L=1$ and $T=1/2$ is written, up to an overall phase, as
\begin{equation}
  Y_{\frac12,1,T_z}(\hat{\bm r})
  =\frac{\bm\tau\cdot\hat{\bm r}}{\sqrt{4\pi}}\chi_{T_z},
  \label{eq:kaon-angular-spinor}
\end{equation}
where $\chi_{T_z}$ is a constant two-component isospinor normalized as
\begin{equation}
\chi_{T_z}^\dagger\chi_{T_z'}=\delta_{T_z T_z'} .
\end{equation}
Thus $(\bm\tau\cdot\hat{\bm r})^2=\bm1_2$ gives $Y^\dagger Y=1/(4\pi)$ for each $T_z$,
which leads to the fact that the contribution to the EMT from the kaon field is spherically symmetric.
The derivative terms also have no angular dependence.
For real $k(r)$,
\begin{align}
  D_iK={}&\frac{e^{-iE_Kt}}{\sqrt{4\pi}}
    \Bigl[k'\hat r_i\,\bm\tau\cdot\hat{\bm r} + \frac{k\cos^2(F/2)}{r}
      \bigl(\tau_i-\hat r_i\bm\tau\cdot\hat{\bm r}\bigr)\Bigr]\chi_{T_z}.
  \label{eq:kaon-angular-covariant-derivative}
\end{align}
Using $\{\tau_a,\tau_b\}=2\delta_{ab}$, one obtains
\begin{align}
  &(D_iK)^\dagger D_jK+(D_jK)^\dagger D_iK\notag\\
  &\quad=\frac{1}{2\pi}\left[
    k'^2\hat r_i\hat r_j
    +\frac{k^2\cos^4(F/2)}{r^2}
       (\delta_{ij}-\hat r_i\hat r_j)\right].
  \label{eq:kaon-angular-derivative-bilinear}
\end{align}
By similarly evaluating the remaining terms in Eqs.~\eqref{eq:kaon-two-derivative-emt} and \eqref{eq:kaon-four-derivative-emt}, we can show that $T_{K}^{00}$ depends only on $r$, while $T_{K}^{ij}$ has precisely the structure of spherically symmetric tensor given in Eq.~\eqref{eq:spatial-emt-densities}.
Thus, from the obtained $T_{K}^{00}$ and $T_{K}^{ij}$, we can extract $\Eden_K$, $p_K$ and $s_K$ for kaons as
\begin{widetext}
\begin{align}
  4\pi\Eden_K[F,k]
    ={}&(1+2c_Ku)|k'|^2
    +\Bigl\{\bigl[1+c_K(F'^2+2u)\bigr]E_K^2
      +m_K^2-m_\pi^2\sin^2\frac F2
      -\frac14(F'^2+2u)+\frac{2w^2}{r^2}\notag\\
    &\qquad+c_K\Bigl[-2u(2F'^2+u)
      +\frac{2w^2}{r^2}(F'^2+4u)\Bigr]\Bigr\}|k|^2
      +\frac{6c_KwF'\sin F}{r^2}(|k|^2)',
      \label{eq:model-kaon-energy}\\
  12\pi p_K[F,k]
    ={}&-(1-2c_Ku)|k'|^2
    +\Bigl\{\bigl[3+c_K(F'^2+2u)\bigr]E_K^2
      -3\Bigl(m_K^2-m_\pi^2\sin^2\frac F2\Bigr)
      +\frac14(F'^2+2u)-\frac{2w^2}{r^2}\notag\\
    &\qquad+c_K\Bigl[-2u(2F'^2+u)
      +\frac{2w^2}{r^2}(F'^2+4u)\Bigr]\Bigr\}|k|^2
      +\frac{6c_KwF'\sin F}{r^2}(|k|^2)',
      \label{eq:model-kaon-trace}\\
  4\pi s_K[F,k]
    ={}&2(1+c_Ku)|k'|^2
    +\Bigl\{-\frac{2w^2}{r^2}-\frac12(F'^2-u)
      +c_K\Bigl[-2(F'^2-u)E_K^2-4u(F'^2-u)
        +\frac{2w^2}{r^2}(F'^2-8u)\Bigr]\Bigr\}|k|^2\notag\\
    &+\frac{6c_KwF'\sin F}{r^2}(|k|^2)'.
      \label{eq:model-kaon-traceless}
\end{align}
where we define
\begin{equation}
  u(r) = \frac{\sin^2 F(r)}{r^2},\qquad w(r)=\cos^2\frac F2,\qquad c_K=\frac{1}{(2ef_\pi)^2}.
  \label{eq:radial-emt-abbreviations}
\end{equation}
for simplicity.
It is noted that the overall factor $4\pi$ in Eqs.~\eqref{eq:model-kaon-energy}-\eqref{eq:model-kaon-traceless} comes from $Y^\dagger Y=1/(4\pi)$.
\end{widetext}

\section{Numerical results}
\label{sec:results}

We first compare the EMT and GFFs with and without the backreaction to the pion field, and then compare the GFFs of the hyperon with those of nucleon. 
We set the parameters as $f_\pi=65.65~\mathrm{MeV}$, $e=4.628$, $m_\pi=138~\mathrm{MeV}$,  $m_K=495~\mathrm{MeV}$ and $\Nc=3$, which are the same as those used in Ref.~\cite{Cebulla:2007ei}.

\subsection{Effect of the backreaction to the pion field}
\label{sec:backreaction-d-ordering}

To elucidate the effect of the backreaction on the pion field, we compare the EMT including the backreaction in Eq.~\eqref{eq:bound-state-emt-nlo} up to ${\mathcal O}(N_c^0)$
\begin{equation}
T_{Y}^{\mu\nu}
    =T_\pi^{\mu\nu}[F_0]+T_{\pi,1}^{\mu\nu}[F_0,F_1] +T_K^{\mu\nu}[F_0,k_0],
\label{EMT: Y final}
\end{equation}
with that obtained in the fixed-background approximation:
\begin{equation}
  T_{Y,\mathrm{fix}}^{\mu\nu}
    =T_\pi^{\mu\nu}[F_0]+T_K^{\mu\nu}[F_0,k_0].
  \label{eq:numerical-fixed-background-emt}
\end{equation}

\begin{figure}[tbp]
  \centering
  \includegraphics[width=\columnwidth]{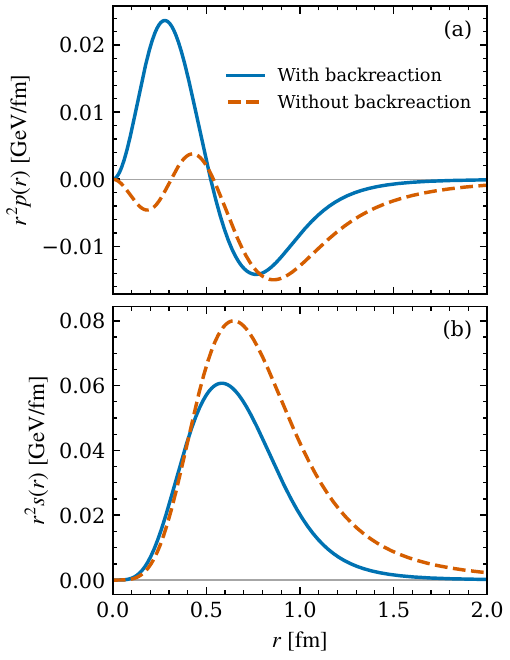}
  \caption{Radial coefficients of the spatial EMT of the hyperon: (a) $r^2p(r)$ and
(b) $r^2s(r)$, both in units of $\mathrm{GeV}/\mathrm{fm}$.
The solid blue and dashed orange curves show the results with and without
pion backreaction, respectively.
  }
  \label{fig:cebulla-backreaction-ps}
\end{figure}

We display the numerical results of $p(r)$ and $s(r)$ for the hyperon obtained by the spatial EMT in Fig.~\ref{fig:cebulla-backreaction-ps}.
As shown in the upper panel of Fig.~\ref{fig:cebulla-backreaction-ps}, when the backreaction is not included, the balance of the positive and negative contributions to the trace part $p(r)$ (orange dashed curve) is substantially broken, which implies that the von Laue condition is not satisfied.
Furthermore, we can confirm that the resultant $p(r)$ and $s(r)$ without backreaction do not satisfy the condition in Eq.~\eqref{eq:local-stress-conservation}, i.e., the EMT in Eq.~\eqref{eq:numerical-fixed-background-emt} is not conserved.
On the other hand, when the backreaction is included, $p(r)$ (blue curve in Fig.~\ref{fig:cebulla-backreaction-ps}) satisfies the von Laue condition.
The resultant $p(r)$ and $s(r)$ satisfy the condition in Eq.~\eqref{eq:local-stress-conservation}, i.e., the EMT in Eq.~(\ref{EMT: Y final}) is actually conserved.

\begin{figure}[tbp]
  \centering
  \includegraphics[width=\columnwidth]{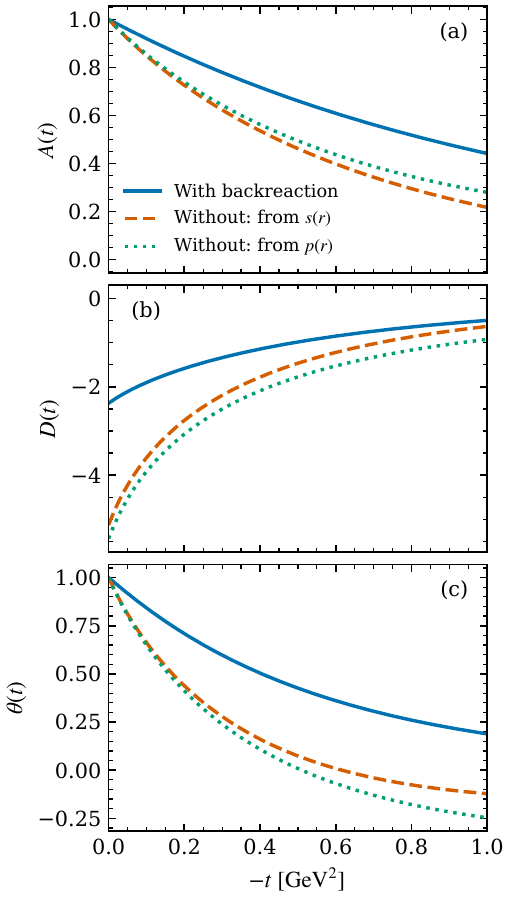}
  \caption{
  Hyperon GFFs with and without the backreaction to the pion field: (a) $A(t)$, (b) $D(t)$, and (c) $\theta(t)$. The solid blue curve indicates the results with the backreaction, while dashed orange and dotted green curves show the results without the backreaction calculated by $s(r)$ and $p(r)$, respectively.
  }
  \label{fig:cebulla-backreaction-gffs}
\end{figure}

Figure~\ref{fig:cebulla-backreaction-gffs} shows the momentum dependence of the hyperon GFFs, with and without the backreaction, and Table~\ref{tab:cebulla-backreaction-gff-radii} lists the corresponding mean-square radii.
Although the mass is unchanged by the backreaction at the order considered, the mean-square radii change significantly.
As shown in Eq.~\eqref{eq:d-from-stress}, there are two ways to compute $D(t)$, $A(t)$ and $\theta(t)$ by using $p(r)$ or $s(r)$. As long as the EMT is conserved, results obtained by the two ways agree with each other.
However, these form factors calculated in the fixed-background approximation show the discrepancy between results obtained from $p(r)$ and $s(r)$.

Furthermore, in the fixed-background approximation, the D-term obtained from $p(r)$ does not agree with that obtained from $s(r)$.
For example, as can be read from Fig.~\ref{fig:cebulla-backreaction-gffs} (b), they are determined as
\begin{align}
  D_{Y,\mathrm{fix}}^{(p)}(0)&=-5.44,\notag\\
  D_{Y,\mathrm{fix}}^{(s)}(0)&=-5.13.
  \label{eq:cebulla-fixed-background-d-moments}
\end{align}
However, including the backreaction term $T_{\pi,1}^{\mu\nu}[F_1]$ restores agreement between the results obtained from $p(r)$ and $s(r)$, and gives
\begin{align}
  D_Y(0)&=-2.37.
  \label{eq:cebulla-with-backreaction-d-moment}
\end{align}
It is noteworthy that, while the mass receives no correction from the backreaction, the D-term differs by approximately a factor of $2$.

\begin{table}[t]
  \caption{Mean-square radii of $A(t)$, $D(t)$, and $\theta(t)$ for the hyperon with and without backreaction on the pion field, defined in Eq.~\eqref{eq:gff-radii}, in $\mathrm{fm}^2$.
  }
  \label{tab:cebulla-backreaction-gff-radii}
  \begin{ruledtabular}
  \begin{tabular}{lrrr}
    Backreaction & $\langle r^2\rangle_A$ & $\langle r^2\rangle_D$ & $\langle r^2\rangle_\theta$ \\
    \hline
    With & 0.19 & 0.81 & 0.39 \\
    Without: $p(r)$ & 0.38 & 0.86 & 1.04 \\
    Without: $s(r)$ & 0.40 & 0.96 & 1.01 \\
  \end{tabular}
  \end{ruledtabular}
\end{table}

\subsection{Nucleon-hyperon comparison}

\begin{figure}[tbp]
  \centering
  \includegraphics[width=\columnwidth]{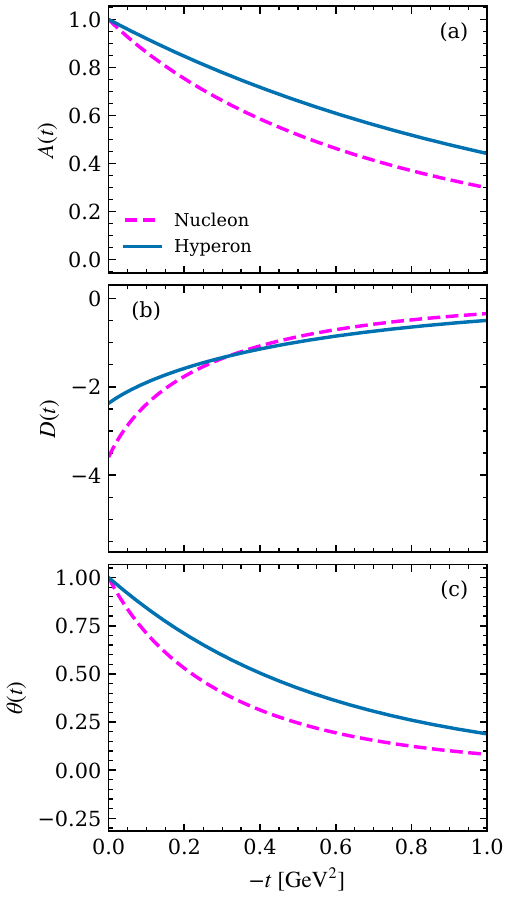}
  \caption{Nucleon (dashed magenta) and hyperon (solid blue) GFFs: (a) $A(t)$, (b) $D(t)$, and (c) $\theta(t)$.
  }
  \label{fig:cebulla-gffs}
\end{figure}

Having established the role of the backreaction, we compare the resultant GFFs of the hyperon with those of the nucleon.
We show the form factors $A(t)$, $D(t)$, and $\theta(t)$ for the nucleon and hyperon in Fig.~\ref{fig:cebulla-gffs}.
This shows that the absolute values of the form factors $A(t)$, $D(t)$, and $\theta(t)$ for hyperon are larger than those for nucleon [$|D(t)|$ for hyperon is slightly smaller for $0.3 \, \mathrm{MeV}<-t$ though], which implies that sizes of them in the coordinate space for nucleon are larger than those for hyperon.
The smallness of the sizes is also seen from the mean-square radii of $A(t)$, $D(t)$, and $\theta(t)$, as shown in Table~\ref{tab:cebulla-gff-radii}.
This indicates a hyperon has a more compact structure than the nucleon.

\begin{table}[t]
  \caption{
  Mean-square radii of the form factors $A(t)$, $D(t)$ and $\theta(t)$ of nucleon and hyperon, defined in Eq.~\eqref{eq:gff-radii}.
  The unit is $\mathrm{fm}^2$.
  }
  \label{tab:cebulla-gff-radii}
  \begin{ruledtabular}
  \begin{tabular}{crrr}
    State & $\langle r^2\rangle_A$ & $\langle r^2\rangle_D$ & $\langle r^2\rangle_\theta$ \\
    \hline
    $N$ & 0.36 & 1.20 & 0.90 \\
    $Y$ & 0.19 & 0.81 & 0.39 \\
  \end{tabular}
  \end{ruledtabular}
\end{table}

\begin{figure}[tbp]
  \centering
  \includegraphics[width=\columnwidth]{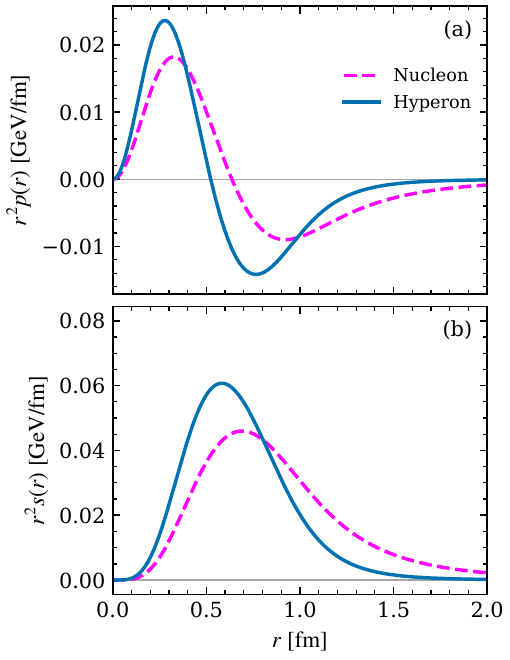}
  \caption{
  Radial coefficients of the spatial EMT for nucleon (dashed magenta) and hyperon (solid blue): (a) $r^2p(r)$ and (b) $r^2s(r)$, in units of $\mathrm{GeV}/\mathrm{fm}$.
  }
  \label{fig:cebulla-nucleon-hyperon-ps}
\end{figure}

Both D-terms are negative and satisfy
\begin{equation}
  |D_Y(0)|<|D_N(0)|.
  \label{eq:numerical-d-hierarchy}
\end{equation}
as shown in Table~\ref{tab:reference-results}.
This hierarchy also shows the compact structure of the hyperon.
We decompose the ratio of the hyperon and nucleon D-terms as follows:
\begin{align}
  \frac{D_Y(0)}{D_N(0)}
    &=\frac{M_Y}{M_N}\,
      \frac{D_Y(0)/M_Y}{D_N(0)/M_N}\notag\\
    &\simeq 1.14 \times 0.58\simeq0.66.
  \label{eq:cebulla-dterm-mass-moment-ratio}
\end{align}
Here $D_Y(0)/M_Y = \int d^{3}r \, r^{2} p_{Y}(r)$ and $D_N(0)/M_N = \int d^{3}r \, r^{2} p_{N}(r)$ are the spatial moments of the hyperon and nucleon, respectively, defined in Eq.~\eqref{eq:d-term}.
As seen in Eq.~\eqref{eq:cebulla-dterm-mass-moment-ratio}, the large mass of the hyperon enhances the magnitude of the D-term. There is an approximate 14\% increase in mass between the nucleon and the hyperon. However, the spatial moment decreases by 42\%, which indicates that the size of the hyperon is smaller than that of the nucleon. Finally, the small D-term of the hyperon is obtained.

\begin{table}[t]
  \caption{Masses and D-terms of the nucleon and the hyperon.}
  \label{tab:reference-results}
  \begin{ruledtabular}
  \begin{tabular}{rrrr}
    $M_N \, [\mathrm{MeV}]$ & $M_Y \, [\mathrm{MeV}]$ & $D_N(0)$ & $D_Y(0)$ \\
    \hline
    1086 & 1240 & -3.59 & -2.37 \\
  \end{tabular}
  \end{ruledtabular}
\end{table}

\begin{table}[htbp]
  \caption{D-terms of nucleon and hyperon for four parameter sets used in Refs~\cite{Adkins:1983ya,Lee:2003eg}, with $m_K=495~\mathrm{MeV}$ and $\Nc=3$ throughout.
  }
  \label{tab:parameter-sets}
  \begin{ruledtabular}
  \begin{tabular}{crrrrr}
    Set & $f_\pi$ [MeV] & $e$ & $m_\pi$ [MeV] & $D_N(0)$ & $D_Y(0)$ \\
    \hline
    1 & 64.5 & 5.45 &   0 & $-2.75$ & $-0.92$ \\
    2 & 93.0 & 4.75 &   0 & $-4.76$ & $-2.60$ \\
    3 & 64.5 & 5.45 & 138 & $-1.97$ & $-1.16$ \\
    4 & 93.0 & 4.75 & 138 & $-3.65$ & $-2.52$ \\
  \end{tabular}
  \end{ruledtabular}
\end{table}

We also check the parameter dependence of the hierarchy of the D-terms in Eq.~\eqref{eq:numerical-d-hierarchy}.
As shown in Table~\ref{tab:parameter-sets}, it remains unchanged across the various parameter sets.

Consistently, the hyperon $r^2p(r)$ and $r^2s(r)$ are enhanced at short distances and suppressed at larger distances relative to those of the nucleon, as shown in Fig.~\ref{fig:cebulla-nucleon-hyperon-ps}.
Then, following the interpretation of the pressure given in Ref.~\cite{Polyakov:2018zvc}, we conclude that the pressure inside the hyperon is larger than that for the nucleon.

\section{Summary and outlook}
\label{sec:conclusions}

We presented a systematic construction of a conserved EMT in the model where the Skyrmion is coupled to additional dynamical fields based on the large-$N_c$ expansion.
The key ingredient of the construction is the backreaction to the pion field which is omitted in the fixed-background approximation. In fact, this contribution to the EMT is of the same order as the ones from the additional fields. Furthermore, we found that including both contributions restores the conservation of the EMT consistently with the large-$\Nc$ expansion. Thus, including both the additional field and the backreaction is necessary to construct the EMT.

We applied the above construction to the model for hyperons based on the CK bound-state approach, and showed that backreaction substantially affects the GFFs.
The resultant function $p(r)$ satisfies the von Laue condition, and it together with $s(r)$ satisfies the consistency condition~\eqref{eq:local-stress-conservation}.
We also demonstrated that in the fixed-background approximation, we could not derive $p(r)$ and $s(r)$ in a consistent way. The predicted form factors computed from $p(r)$ and $s(r)$ in the fixed-background approximations do not agree with each other, while the form factors should be consistent as long as the EMT is conserved.

We compared the GFFs of the hyperon with those of the nucleon. 
Our results showed that the magnitude of D-term for the hyperon GFF is smaller than that for the nucleon GFF.
It was also shown that the mean square radii of three form factors $A(t)$, $D(t)$ and $\theta(t)$ are smaller for the hyperon, consistently with the greater concentration of its radial EMT profiles $p(r)$ and $s(r)$ at short distances.
These results imply a more compact structure for the hyperon EMT.

We would like to note that, while the more compact structure of the hyperon relative to the nucleon, as inferred from $A(t)$, is consistent with the results of Refs.~\cite{Won:2022cyy,Liu:2025qja}, the hierarchy of the D-term differs from that obtained in the SU(3) chiral quark-soliton model~\cite{Won:2022cyy}.
This might indicate that we need further comparisons between models and with lattice QCD.

Future directions include collective quantization with the associated backreaction, higher-order corrections, and extensions incorporating scalar fields and vector mesons.
Scalar fields would allow a more explicit treatment of the scale anomaly and its relation to the trace form factor $\theta(t)$.
Applying the construction to the bound-state approach for the heavy-quark sector would also provide a route to the GFFs of singly heavy baryons.

\begin{acknowledgments}
We thank Mamiya Kawaguchi, Daisuke Fujii, and Ho-Yeon Won for helpful discussions.
The work of M.T. is supported by ``THERS Make New Standards Program for the Next Generation Researchers" and JST SPRING, Grant Number JPMJSP2125.
This work is supported in part by JSPS KAKENHI Grant Nos.~23H05439 and 24K07045 (M.H.), and 26K07104 (Y.Y.).
\end{acknowledgments}

\bibliography{references}

\end{document}